\documentclass[preprint,12pt]{elsarticle}

\usepackage{amssymb}
\usepackage{mathrsfs}
\usepackage{calligra}
\usepackage{color}
\usepackage{graphicx}

\newcommand{\be}{\begin{equation}}
\newcommand{\ee}{\end{equation}}
\newcommand{\bea}{\begin{eqnarray}}
\newcommand{\eea}{\end{eqnarray}}

\journal{Annals of Physics}

\begin{document}

\begin{frontmatter}



\title{Duality and topology}


\author{P. D. Sacramento and V. R. Vieira}
\address{\textit CeFEMA,
Instituto Superior T\'ecnico, Universidade de Lisboa, Av. Rovisco Pais, 1049-001 Lisboa, Portugal and}

\address{
Beijing Computational Science Research Center, Beijing 100084, China}

\begin{abstract}
Mappings between models may be obtained by unitary transformations with
preservation of the spectra but in general a change in the states.
Non-canonical transformations in general also change the statistics
of the operators involved.
In these cases one may expect a change of topological properties as a consequence
of the mapping. Here we consider some dualities resulting from mappings, 
by systematically using a Majorana fermion representation of spin and
fermionic problems. We focus on the change of topological invariants that results from
unitary transformations taking as examples the mapping between a
spin system and a topological superconductor, and between different fermionic systems.
\end{abstract}


\end{frontmatter}


\section{Introduction}

Many-body interacting systems are problems that are hard to solve, often
requiring non-perturbative approaches to properly describe
their cooperative phenomena. In general the complexity of the problem
may be reduced identifying the dominant modes that govern the behavior
of the system, particularly when a low-energy description is enough.
Also, many authors have considered various transformations
between variables or operators (depending if the system is classical
or quantum, respectively) to obtain a good description of the
system behavior and, in some cases, exactly solve the problem. 

Depending on the problem, transformations from one set of
operators to another may involve canonical transformations \cite{mele},
preserving the statistics, or non-canonical transformations,
where the statistics is altered. Typical examples are bosonization
of a fermionic problem or, reversely, fermionization. Also, often it is
convenient to transform between spin and fermionic
problems. Some transformations are exact \cite{ostlund,kumar,angelucci} but in some cases there is
an enlargement of the Hilbert space, and a projection to the physical
subspace is required \cite{barnes,coleman0,revlee,millis,kotliar,dorin,kane,arovas}. 
The simplest cases involve local transformations from
one set of operators to another, but non-local transformations
are also convenient in some cases. Also, some transformations 
have an intrinsic non-linear character \cite{mele}. A familiar example are
the bilinear representations of spin operators in terms of bosonic
or fermionic operators \cite{auerbach}.
Since the transformation between the two sets of operators is bilinear,
it naturally provides a way for a non-canonical transformation with
consequent change of statistics. 
In some cases it has been shown that performing a mapping between
spin and fermionic systems it is possible to reduce an apparent interacting term
into a free problem, with the consequent exact solution. An example is provided
by a X-Y chain which may be reduced to a problem of free (in the sense of
quadratic) problem of spinless fermions \cite{lieb61,springer}, the so-called fermionic 
Kitaev model \cite{kitaev}. This example illustrates a problem
of statistical transmutation of an original problem of spins into a
problem of fermions. This is well known to be achieved applying a Jordan-Wigner
transformation \cite{jordan} which traditionally is understood as a transformation between
operators defined such that their commutation relations are satisfied. 
Other non-canonical transformations include the Schrieffer-Wolff transformation \cite{wolff}
between the Anderson model and the Kondo model and bosonization between fermions
and a bosonic field \cite{bosonization}.
The Jordan-Wigner transformation that leads to the Kitaev model is reviewed in Appendix A.

In this work we will focus on fermionic and spin-$1/2$ systems.
Among the many representations for spin systems one is particularly
convenient since it allows an exact preservation of the commutation
relations of the spin operators \cite{berezin,vrv1,us1}. There is some enlargement of the
Hilbert space but it just leads to a multiplying factor in the partition
function of the system \cite{tsvelik1,coleman,shastry}. Specifically, the spin operators may be represented
by a bilinear representation in terms of three Majorana operators.
The spin-$1/2$ representation in terms of Majoranas has been applied in several
contexts \cite{vrv1,us1,shastry,coleman,vrv2,turkowski}.
Majorana operators may also be used to represent a fermionic operator
in a simple way. A fermionic operator may be understood as containing
real and imaginary parts if these are chosen as hermitian operators, which
is the characteristic property of a Majorana fermion. 
It is therefore convenient to look for transformations between spin and
fermionic operators using a language of transformation operators in terms
of Majorana fermions. Looking for general transformations
using Majorana operators, it is possible to consider different choices of
representations which enable both canonical and non-canonical transformations.
In addition, it has been
proposed that they provide a convenient way to understand the general properties
of transformations between operators \cite{nilsson3,thesis}. 

The interest in Majorana fermions has recently been revived due to their possible
relevance in quantum computation problems
and have been proposed to be observed in the context of topological superconductors
\cite{alicea}.
Even though appropriate materials are difficult to find in nature, several
engineered possibilities have been proposed such as semiconductor wires with
strong spin-orbit coupling placed on top of a conventional superconductor
and in the presence of a Zeeman field \cite{wire}, or magnetic impurities on top of
a conventional superconductor \cite{bernevig}. In these systems the Majorana modes are associated
with edge modes at the border between a topologically non-trivial system and
a trivial system. 

The transformation between sets of operators or variables leads to an
equivalent problem whose
Hamiltonian is expressed in terms of other physical quantities.
In some loose sense we may think of a transformation as a relation between
dual Hamiltonians. Dualities appear in physics in various contexts from dualities in
classical electromagnetism (between electric and magnetic fields at the level
of Maxwell's equations), dualities in statistical physics problems (such as
the Kramers-Wannier identities \cite{kramers} that relate the partition function of the
system in the low and high temperature limits) and dual lattices (by establishing a
relation between variables or operators defined at the corners or links of some
lattice problem). 

As pointed out recently \cite{cobanera1} the duality transformations do not need to relate strong and
weak coupling regimes (although they are particularly useful when this occurs),
and usually are non-local relationships involving often some kind of extended strings of
operators. 
Intrinsic to the idea of duality and the search of an alternative set of operators
to describe the properties of some Hamiltonian, is the equivalence between the two
descriptions. An expected minimal requirement is that the spectrum of the Hamiltonian
is preserved. An obvious way to achieve this is to consider unitary transformations
since the spectrum is preserved. However, in general, the states are not preserved
even though the relation between them is determined by the choice of the
unitary operator. As a consequence of the change of states it has been
pointed out that, in general, level degeneracy changes and therefore the
correspondence between the two Hamiltonians is not complete. 

It has also been pointed out recently that the usual non-local character of the
duality transformation suggests that the use of bond operators may be more
convenient than local relations. In reference \cite{cobanera1} several transformations have
been considered and the standard example of a Jordan-Wigner transformation
between a spin-$1/2$ problem and a spinless fermion non-interacting 
Hamiltonian has been derived with the bond-duality approach.
In the case of nearest-neighbor couplings this transformation has been known for
a long time to fully diagonalize the Hamiltonian, since in the fermionic language
is non-interacting.  
However, it is also known that the Jordan-Wigner transformation may be seen
as the result of a local unitary transformation \cite{coleman} (understood as a product of
local transformations across all lattice sites). In this work we will
follow a similar procedure and therefore consider a less stringent definition of duality
(see also, for instance \cite{huang}).

The mapping between the $XY$ spin model in a transverse field and the Kitaev
superconducting model \cite{kitaev} reveals another interesting result. While the original
spin problem is topologically trivial, the resulting transformed Hamiltonian
has topological regimes. Therefore, as a consequence of the exact transformation
between the two problems, it appears that the topological properties have changed.
On the other hand, it has been shown recently that it is possible to transform
topological insulators to topological superconductors \cite{cobanera3}. In this case this is a canonical
transformation. It was shown that topological phases are matched to topological
phases and trivial phases to trivial phases. This suggests that a non-canonical
transformation may be required to change topology. 

Topological systems appear in various contexts. To name a few, among those that
have received recently considerable attention are the topological insulators
and topological superconductors \cite{hasan,zhang,hanson} due to their robust edge states with possible
applications in dissipation free transport and quantum computation.
Other well-known topological systems are some spin chains.
An example of phases with topological origin are the gapped Haldane phases \cite{haldane}
of odd integer spin chains that
find an exact realization in the
gapped AKLT phases \cite{aklt}, where non-local string order has been found. This hidden order 
is the result of a hidden order symmetry \cite{nijs,tasaki} 
and may be understood as the result of a 
non-local unitary transformation \cite{oshikawa1,oshikawa2}.
One may search for
correlation functions in a topological system that are related via duality with the order parameters 
in the trivial but ordered dual system \cite{cobanera2,nussinov,tee}.
True topologically ordered systems display long-range topological order that may not be
eliminated by any local transformation \cite{wen,read}. 

The work is organized as follows:
In section 2 we consider the action of general unitary transformations on fermion and
spin operators, expressed in terms of Majorana fermion operators. 
In section 3 we consider mappings between spin and fermionic systems
using a non-canonical unitary transformation that allows a mapping between a nearest-neighbor
spin problem and a system of free spinless fermions. 
In section 4 we consider topological invariants for different representations
of the fermionic operators of Kitaev's model. 
In section 5 we focus on the transformation between a spin problem
and a corresponding fermionic problem and discuss the role of boundary conditions
on topological properties.
We conclude with section 6.

\section{Unitary transformations of fermion and spin operators}

\subsection{Fermion, Majorana and spin operators}

In general, a fermion operator at site $j$ may be written 
in terms of two hermitian operators, $\gamma_1, \gamma_2$, in the
following way
\bea
c_{j,\sigma} &=& \frac{1}{2} \left( \gamma_{j, \sigma, 1} + i \gamma_{j, \sigma, 2} \right) \nonumber \\
c_{j,\sigma}^{\dagger} &=& \frac{1}{2} \left( \gamma_{j, \sigma, 1} -
i \gamma_{j, \sigma, 2} \right)
\eea
The index $\sigma$ represents internal degrees of freedom of the fermionic operator, such as spin
and/or sublattice index.
In general, the $\gamma$ operators we will consider are hermitian and satisfy a Clifford algebra
$\{\gamma_{m,i},\gamma_{n,j} \}=2 \delta_{ij} \delta_{mn} I$,
where $m,n=j,\sigma$, and $I$ is the identity. We use the normalization $\gamma_i^2=1$ and $i=1,2,3$.

Majoranas also allow a representation of spin-$1/2$ problems.
One needs three Majorana operators to
represent local spin operators as 
\be
S_x = -(i/2) \gamma_2 \gamma_3; \hspace{1cm}
S_y = -(i/2) \gamma_3 \gamma_1; \hspace{1cm}
S_z = -(i/2) \gamma_1 \gamma_2.
\label{spinsops}
\ee
It is convenient to define the operators
$\mathscr{S}_i=\frac{-i}{2}\epsilon_{ijk}\gamma_j\gamma_k$, differing in normalization from
the standard spin operators $\vec{S}=\frac{1}{2}\vec{\mathscr{S}}$, and the operator
$I_3=-i\gamma_1\gamma_2\gamma_3=\frac{-i}{6}\epsilon_{ijk}\gamma_i\gamma_j\gamma_k$.

With the definition of these operators, the multiplication rule of the Majorana
operators becomes
\be
\gamma_i\gamma_j=\delta_{ij}I+i\epsilon_{ijk}\mathscr{S}_k
\ee
from which it follows that $\vec{\gamma}^2=3I$.
The multiplication rules of the Majorana and spin operators are given by 
\begin{eqnarray}
\gamma_i\mathscr{S}_j&=&\delta_{ij}I_3+i\epsilon_{ijk}\gamma_k \\
\mathscr{S}_i\gamma_j&=&\delta_{ij}I_3+i\epsilon_{ijk}\gamma_k
\end{eqnarray}
and
\be
\mathscr{S}_i\mathscr{S}_j=\delta_{ij}I+i\epsilon_{ijk}\mathscr{S}_k
\ee
together with $I_3^2=I$, $I_3\gamma_i=\gamma_i I_3=\mathscr{S}_i$, and 
$I_3\mathscr{S}_i=\mathscr{S}_i I_3=\gamma_i$. One also has
$\vec{\mathscr{S}}^2=3I$ and
$\vec{\gamma}\cdot\vec{\mathscr{S}}=\vec{\mathscr{S}}\cdot\vec{\gamma}=3I_3$.
These equations can be summarized as that the operators $\gamma_i$ and
$\mathscr{S}_i$ follow multiplication rules similar to the Pauli matrices, but
paying attention to their nature, i.e. being odd or even in the Majorana
operators, and that the multiplication by $I_3$ transforms $\gamma_i$ into
$\mathscr{S}_i$ and vice versa.

\subsection{Enlargement of Hilbert space}

An interesting question when representing fermions and spins using
Clifford algebras is the enlargement of the number of states and the resulting
degeneracy of the states.
In the case of the treatment of fermion systems there is no enlargement of the 
number of states. With the fermionic operators 
$c = \frac{1}{2} \left(\gamma_1 + i \gamma_2 \right),
c^{\dagger} = \frac{1}{2} \left(\gamma_1 - i \gamma_2 \right)$ 
one can define the Majorana operators $\gamma_1 = c^{\dagger}+c$ and
$\gamma_2 = i \left( c^{\dagger}-c\right)$. The operator 
$\mathscr{S}_3 = -i \gamma_1 \gamma_2 = \left(2 c^{\dagger} c -1 \right)$
anticommutes with both 
$\gamma_1$ and $\gamma_2$, but is not independent from them, and there is no
enlargement of the number of states, which continues to be $2$.
However, when one considers bilinear representations of spin operators in terms
of Clifford operators an enlargement of the number of states occurs, resulting
from the $\mathbb{Z}_2$ symmetry implied by the bilinear representation.

In the case of a single spin operator, there is a duplication of the number of 
states, with $4$ states instead of $2$. The easier way to understand this is to 
introduce an extra Clifford operator to pair with the third to give a 
second pair of creation and destruction operators. Applying this procedure to a
set of $N$ spins $\frac{1}{2}$ one has then $2^{2N}$ states instead of $2^N$.
In general, if one has $N$ spins $\frac{1}{2}$, the number of states is $2^N$.
Introducing three Clifford operators for each spin, one has $3N$ Majorana
operators. If the number of sites $N$ is even, we can associate them in pairs
obtaining $\frac{3N}{2}$ pairs of fermion operators, and the number of states 
will be $2^\frac{3N}{2}$. If the number of states is odd, one introduces then 
an extra Clifford operator and the number of states is $2^\frac{3N+1}{2}$.

As is well known, the complex Clifford algebras have the algebraic isomorphisms
$Cl(n)=\mathbb{C}(2^\frac{n}{2})$ if $n$ is even, and 
$Cl(n)=\mathbb{C}(2^\frac{n-1}{2})\oplus \mathbb{C}(2^\frac{n-1}{2})$
if $n$ is odd \cite{brauer,ryu}.
If $n$ is even, the algebra is central simple, but if $n$ is odd, besides the
identity, the center also includes the product of all the Clifford operators,
and one can define a projection into an even and an odd algebra 
under the normalized product of all the Clifford operators, similarly to the results shown in
Appendix B for 
$I_\pm=\frac{1}{2}(I\pm I_3)$,
$\Sigma^\pm_i=\frac{1}{2}(\mathscr{S}_i\pm\gamma_i)$
and $I_3$.

The case of two spins and of its degeneracy has been discussed in the literature
\cite{coleman,shastry}, showing that for the Heisenberg interaction, the singlet
and triplet states become both doubly represented, and that, as a result, the
partition function is simply multiplied by this multiplicity factor. In
\cite{coleman}, the case of an electron with spin is also discussed at length,
showing that the usual spin electron operator, associated to the spin degree of
freedom, and the Nambu pseudospin operator, associated to the charge degree of
freedom, are orthogonal, with their sum providing a doubly representation of
$SU(2)$. Again, the partition function is simply multiplied by a multiplicity
factor, even in the case of many electrons with a spin-spin interaction between
the sum of the spin and pseudospin of each electron.

\subsection{Action of unitary and hermitian transformations on local operators}

Let us first consider general local transformations on the spin and Majorana operators 
of the type $U^2=I$, therefore unitary
transformations that are also hermitian operators.
Let us look for solutions of the type
\be
U=z I + \omega I_3 + \vec{x} \cdot \vec{\gamma} + \vec{y} \cdot \vec{\mathscr{S}}
\ee
Here $I_3=i \gamma_1 \gamma_2 \gamma_3$, 
$\vec{\mathscr{S}}=2 \vec{S}$ and $z,\omega,\vec{x},\vec{y}$ are real numbers.
These imply that $U$ is unitary. The solutions of $U^2=I$ may be organized into the following
classes: 
\begin{itemize}

\item (i) $U=I$

\item (ii) $U=I_3$

\item (iii) $U=\frac{1}{2} \left( I + I_3 + \vec{n}_- \cdot \frac{1}{2} \left(
\vec{\mathscr{S}} -\vec{\gamma} \right) \right)$, with $|\vec{n}_- |=1$

\item (iv) $U = \pm \frac{1}{2} \left( I-I_3 + \vec{n}_+ \cdot \frac{1}{2} \left(
\vec{\mathscr{S}} +\vec{\gamma} \right) \right)$, with $|\vec{n}_+ |=1$

\item (v) $U=\cos \frac{\theta}{2} \vec{n}_1 \cdot \vec{\gamma} + \sin 
\frac{\theta}{2} \vec{n}_2 \cdot
\vec{\mathscr{S}}$, with
$|\vec{n}_1| = |\vec{n}_2 |=1$ and $\vec{n}_1 \cdot \vec{n}_2 =0$.

\end{itemize}

Both in classes (i) and (ii) we find that
acting on operators on a given site we get
$U \vec{\gamma} U^{\dagger} = \vec{\gamma}$ and
$U \vec{\mathscr{S}} U^{\dagger} = \vec{\mathscr{S}} $.
In the case of class (iii) we get
\bea
U \frac{1}{2} \left( \vec{\mathscr{S}} -\vec{\gamma} \right) U^{\dagger}
&=& 2 \vec{n} \left( \vec{n} \cdot \frac{1}{2} (\vec{\mathscr{S}} -\vec{\gamma} ) \right)
-\frac{1}{2} \left( \vec{\mathscr{S}} -\vec{\gamma} \right)
\nonumber \\
U \vec{\gamma} U^{\dagger} &=& \vec{\mathscr{S}} - \vec{n} \left( \vec{n} \cdot
(\vec{\mathscr{S}} -\vec{\gamma} ) \right) \nonumber \\
U \vec{\mathscr{S}} U^{\dagger} &=& \vec{\gamma} + \vec{n} \left( \vec{n} \cdot
(\vec{\mathscr{S}}-\vec{\gamma} ) \right) 
\eea
In the case of class (iv) we get that
\bea
U \vec{\gamma} U^{\dagger} &=& -\vec{\mathscr{S}} + \vec{n} \left( \vec{n} \cdot
(\vec{\mathscr{S}}+\vec{\gamma} ) \right) \nonumber \\
U \vec{\mathscr{S}} U^{\dagger} &=& -\vec{\gamma} + \vec{n} \left( \vec{n} \cdot
(\vec{\mathscr{S}} +\vec{\gamma} ) \right) 
\eea
Finally, in the case of class (v) we get that
\bea
U \vec{\gamma} U^{\dagger} &=& -\vec{\gamma} + 2\left[ (\cos \frac{\theta}{2})^2 \vec{n}_1 \left(
\vec{n}_1 \cdot \vec{\gamma} \right) \right. \nonumber \\
&+& \left. (\sin \frac{\theta}{2})^2 \vec{n}_2 \left( \vec{n}_2 \cdot \vec{\gamma} \right) \right. \nonumber \\
&+& \left. \sin \frac{\theta}{2} \cos \frac{\theta}{2}
\left( \vec{n}_1 (\vec{n}_2 \cdot \vec{\gamma} ) + \vec{n}_2 (
\vec{n}_1 \cdot \vec{\gamma}) \right) \right] \nonumber \\
U \vec{\mathscr{S}} U^{\dagger} &=& -\vec{\mathscr{S}} + 2\left[ (\cos \frac{\theta}{2})^2 \vec{n}_1 \left(
\vec{n}_1 \cdot \vec{\mathscr{S}} \right) \right. \nonumber \\
&+& \left. (\sin \frac{\theta}{2})^2 \vec{n}_2 \left( \vec{n}_2 \cdot \vec{\mathscr{S}} \right) \right. \nonumber \\
&+& \left. \sin \frac{\theta}{2} \cos \frac{\theta}{2}
\left( \vec{n}_1 (\vec{n}_2 \cdot \vec{\gamma} ) + \vec{n}_2 (
\vec{n}_1 \cdot \vec{\gamma}) \right) \right]
\eea

\subsection{General unitary operations}

Let us now consider local transformations that are unitary but not hermitian.
Some possible examples are illustrated next.

(a) $U=\frac{1}{\sqrt{2}} \left( 1+i \vec{n} \cdot \vec{\mathscr{S}} \right)$, with
$|\vec{n}|=1$. The action of this operator leads to
\bea
U \vec{\mathscr{S}} U^{\dagger} &=& \vec{n} \left( \vec{n} \cdot \vec{\mathscr{S}} \right)
+\vec{n} \times \vec{\mathscr{S}} \nonumber \\
U \vec{\gamma} U^{\dagger} &=& \vec{n} \left( \vec{n} \cdot \vec{\gamma} \right)
+\vec{n} \times \vec{\gamma} 
\eea
The action of this unitary operator does not change the nature of the operators and therefore
is an example of a canonical transformation.

(b) $U=\frac{1}{\sqrt{2}} \left( 1+i \vec{n} \cdot \vec{\gamma} \right)$, with
$|\vec{n}|=1$. The action of this operator leads to
\bea
U \vec{\gamma} U^{\dagger} &=& \vec{n} \left( \vec{n} \cdot \vec{\gamma} \right)
+\vec{n} \times \vec{\mathscr{S}} \nonumber \\
U \vec{\mathscr{S}} U^{\dagger} &=& 
\vec{n} \left( \vec{n} \cdot \vec{\mathscr{S}} \right)
+\vec{n} \times \vec{\gamma} 
\eea
mixing the nature of the operators and therefore
is an example of a non-canonical transformation.

(c) $U=\frac{1}{\sqrt{2}} \left( I_3 +i \vec{n} \cdot \vec{\gamma} \right)$, with
$|\vec{n}|=1$ and $I_3=-i \gamma_1 \gamma_2 \gamma_3$. The action of this operator leads to
a set of canonical transformations
\bea
U \vec{\gamma} U^{\dagger} &=& \vec{n} \left( \vec{n} \cdot \vec{\gamma} \right)
+\vec{n} \times \vec{\gamma} \nonumber \\
U \vec{\mathscr{S}} U^{\dagger} &=& \vec{n} \left( \vec{n} \cdot \vec{\mathscr{S}} \right)
+\vec{n} \times \vec{\mathscr{S}} 
\eea

(d) $U= \cos \theta/2 + i \sin \theta/2 \vec{n} \cdot \vec{\gamma}$ leads to
\bea
U \vec{\gamma} U^{\dagger} &=& 
\vec{n} \left( \vec{n} \cdot \vec{\gamma} \right) + \cos \theta \left( \vec{\gamma} -\vec{n} 
\left( \vec{n} \cdot \vec{\gamma} \right)
\right) \nonumber \\
&+& \sin \theta  \left( \vec{n} \times \vec{\mathscr{S}} \right)
\nonumber \\
U \vec{\mathscr{S}} U^{\dagger} &=& 
\vec{n} \left( \vec{n} \cdot \vec{\mathscr{S}} \right) + \cos \theta \left( \vec{\mathscr{S}} -\vec{n} 
\left( \vec{n} \cdot \vec{\mathscr{S}} \right)
\right) \nonumber \\
&+& \sin \theta  \left( \vec{n} \times \vec{\gamma} \right)
\nonumber \\
\eea
which is also non-canonical since it mixes the two types of operators, Majoranas and spin operators.

Note that from
\be
U=\frac{1}{2} \left[ I + I_3 -\vec{n} \cdot \left( \vec{\mathscr{S}}-\vec{\gamma} \right) \right]
\ee
choosing $\vec{n}=\vec{e}_z$ we get that
\bea
U &=& \frac{1}{2} \left[ I+I_3+\gamma_3 -\mathscr{S}_3 \right] \nonumber \\
&=& \frac{1}{2} \left[1-i \gamma_1 \gamma_2 \gamma_3 + \gamma_3 +i \gamma_1 \gamma_2 \right] = U_z
\eea
Also, note that
\be
U_z \mathscr{S}_z U_z = \mathscr{S}_z \hspace{1cm}
U_z \mathscr{S}_x U_z = \gamma_1 \hspace{1cm}
U_z \mathscr{S}_y U_z = \gamma_2
\ee
This class of transformations given by $\vec{n}=\vec{e}_{\alpha}$ leaves one of the spin operator
components invariant,
$U_{\alpha} \mathscr{S}_{\alpha} U_{\alpha} = \mathscr{S}_{\alpha}$
for $\alpha=x,y,z$ $(1,2,3)$. The transformation acts in the perpendicular plane.

e) Consider now the following unitary operator of class (v)
\be
U_v=\cos \frac{\theta}{2} \vec{n}_1 \cdot \vec{\gamma} + \sin \frac{\theta}{2} \vec{n}_2 \cdot \vec{\mathscr{S}}
= U_v^{\dagger}
\ee
with $\vec{n}_1 \cdot \vec{n}_2 =0$. Take for instance $\vec{n}_1=\vec{e}_x, \vec{n}_2=\vec{e}_y$.
Then
\be
U_v=\cos \frac{\theta}{2} \gamma_1 + \sin \frac{\theta}{2} \mathscr{S}_2 = 
\cos \frac{\theta}{2} \gamma_1 -i \sin \frac{\theta}{2} \gamma_3 \gamma_1
\ee
The action of the operator on the spin components is
\bea
U_v \mathscr{S}_x U_v &=& \cos \theta \mathscr{S}_1 + \sin \theta \gamma_2 
\nonumber \\
U_v \mathscr{S}_y U_v &=& -\cos \theta \mathscr{S}_2 + \sin \theta \gamma_1 
\eea
Also, taking two sites $l$ and $j$
\be
U_{v,l} \gamma_{\alpha,j} U_{v,l} = \left( -\cos \theta +i \sin \theta \gamma_{3,l} \right) \gamma_{\alpha,j}
\ee
with $\alpha=1,2,3$.

Having established several possible transformations between spin and fermionic
operators one may now use them to construct exact mappings between 
different models. Our focus here will be on the transformation mentioned above
between a spin model and its fermionic description and on the effect of different
fermionic representations on topology.

\section{Mapping between spins and fermions}

\subsection{Non-canonical unitary transformation}

The Jordan-Wigner transformation may also be constructed (see appendix in Ref. \cite{coleman}) introducing the
local unitary transformation $U_z$. To simplify let us consider the XX model
($J_x=J_y=1$) or the fully anisotropic $X$ model ($J_x=1,J_y=0$) model
\be
H_X = 
 \frac{1}{4} \sum_j \left( S^+_j S^-_{j+1} +  S^-_j S^+_{j+1} 
 + S^+_j S^+_{j+1} +  S^-_j S^-_{j+1} \right) 
\label{xmodel}
\ee
The spin operators may be represented by the Majorana operators as \cite{berezin,vrv1}
$S^+= \gamma_3 \left( \gamma_1 + i \gamma_2 \right)/2$,
$S^-= \left( \gamma_1 - i \gamma_2 \right) \gamma_3 /2$.
Define now usual fermionic operators (non-hermitian) as
$g = \frac{1}{2} \left( \gamma_1 -i \gamma_2 \right)$,
$g^{\dagger} = \frac{1}{2} \left( \gamma_1 + i \gamma_2 \right)$.
We get that
$\gamma_1 = g+g^{\dagger}$ and 
$i \gamma_2 = g^{\dagger}-g$.
Therefore we may write the Hamiltonians as
\bea
H_{XX} &=& \frac{1}{2} \sum_j \left( g_j^{\dagger} g_{j+1} + g_j g_{j+1}^{\dagger} \right)
\gamma_{3,j} \gamma_{3,j+1} \nonumber \\
H_{X} &=& \frac{1}{4} \sum_j \left( g_j^{\dagger} g_{j+1} + g_j g_{j+1}^{\dagger} 
-g_j^{\dagger} g_{j+1}^{\dagger} -g_j g_{j+1}
\right) 
\gamma_{3,j} \gamma_{3,j+1} 
\label{xmodel3}
\eea

The Hamiltonian in terms
of the Majorana and regular fermions is interacting, as evidenced by the
quartic terms in the Hamiltonian. 
A possible way to diagonalize the Hamiltonian is to perform a unitary transformation
that eliminates the $\gamma_3$ operators. This can be achieved using the local unitary and
hermitian operator \cite{coleman}
\bea
U_{z,j} &=& \left(1- g_j^{\dagger} g_j \right) +\gamma_{3,j} g_j^{\dagger} g_j
\nonumber \\
&=& \frac{1}{2} \left[ 1+ i \gamma_{1,j} \gamma_{2,j} + \gamma_{3,j} -i \gamma_{1,j}
\gamma_{2,j} \gamma_{3,j} \right]
\eea
Defining now an operator that is the product over all sites in the one-dimensional system
\be
U_z=\prod_{j=1}^N U_{z,j} = U_{z,1} \cdots U_{z,N}
\ee
we get
\be
U_z \gamma_{3,j} g_j^{\dagger} U_z^{\dagger}=\left(-1 \right)^{\sum_{l=1}^{j-1} n_l}
g_j^{\dagger} 
\label{stringy}
\ee
and its hermitian conjugate. We see that this unitary transformation gives origin to the so-called strings associated 
with the occupation of states to the left of a given lattice site, $j$.
This is like in the Jordan-Wigner transformation. Applying to the Hamiltonian
\be
U_z H_{XX} U_z^{\dagger} = 
-\frac{1}{2} \sum_j \left[ g_j^{\dagger} g_{j+1} + g_{j+1}^{\dagger} g_j \right]
\ee
Similarly \cite{coleman}
\be
U_z H_X U_z^{\dagger} = 
 -\sum_j \left[ g_j^{\dagger} g_{j+1} + g_{j+1}^{\dagger} g_j + g_j^{\dagger} g_{j+1}^{\dagger}
+ g_{j+1} g_j \right]
\ee
which is Kitaev's model \cite{kitaev} for $t=1,\Delta=1$.
In contrast to the original model which described interacting spins
on a lattice, this model describes spinless fermions that may hopp
on a lattice and have a nearest-neighbor $p$-type superconducting pairing, as shown
by the appearance of creation and destruction pair operators.
A magnetic field term is invariant (see Eq. (\ref{spinsops})).
We can also write
\be
H_X = \frac{1}{4} \sum_j \gamma_{2,j} \gamma_{2,j+1} \gamma_{3,j} \gamma_{3,j+1}
\ee
This quartic Hamiltonian is transformed to a quadratic Hamiltonian
by eliminating the $\gamma_3$ terms since the action of the unitary operator $U_{z,j}$ is
of the form
\be
U_{z,j} \gamma_{2,j} \gamma_{3,j} U_{z,j}^{\dagger} = i \gamma_{1,j}
\ee
can be seen as transforming a spin-like operator (bilinear in the Majoranas) into
a Majorana operator. We may also see that
$ U_{z,j} \gamma_{1,j} \gamma_{3,j} U_{z,j}^{\dagger}  = -i \gamma_{2,j}$.
Noting now that
\be
\left( \prod_{l=1}^{j-1} U_{z,l} \right) \gamma_{1,j} \left( \prod_{l=j-1}^1 U_{z,l}^{\dagger}
\right) = \gamma_{1,j} \prod_{l=1}^{j-1} \left[ i \gamma_{1,l} \gamma_{2,l} \right]
\ee
we get
\be
U_z H_X U_z^{\dagger} 
= \frac{i}{4} \sum_j \gamma_{2,j} \gamma_{1,j+1}.
\ee
which is a non-interacting fermionic problem.

\subsection{Relation between fermionic states and spin states and
order parameter correspondence}

The mapping of Kitaev's model to the spin system identifies $\Delta=t$ with $J_y=0$ (see Eq. (\ref{eqa5}).
Consider therefore the $X$ chain in the ferromagnetic case
$H_X = -\sum_j S_{x,j} S_{x,j+1}$.
The groundstate is
$|GS \rangle = |\sigma_x=1; \sigma_x=1; \cdots ; \sigma_x=1; \cdots \rangle $
or $\sigma_x=-1$ at every site ($Z_2$ degeneracy).
The solution for the antiferromagnetic model is presented in \cite{lieb61}.
Recall that $\sigma_x=\mathscr{S}_x= -i \gamma_2 \gamma_3$. Therefore
the action of an operator at a given lattice site on the groundstate is
$(-i) \gamma_2 \gamma_3 |GS \rangle = |GS\rangle $.

Define now new local fermionic operators in terms of the Majorana operators
used to represent the spin operators as
$f_j = \frac{1}{2} \left( \gamma_{2,j} +i \gamma_{3,j} \right) $ and
$f_j^{\dagger} = \frac{1}{2} \left( \gamma_{2,j} -i \gamma_{3,j} \right) $.
Similarly to previous results it is easy to see that
$2 f_j^{\dagger} f_j-1 = i \gamma_{2,j} \gamma_{3,j}$.
Therefore,
$\mathscr{S}_{1,j} = -i \gamma_{2,j} \gamma_{3,j} = -\left( 2 f_j^{\dagger} f_j-1 \right)$.
Therefore we can identify
$|\mathscr{S}_x=1\rangle = |0\rangle_f $,
$|\mathscr{S}_x=-1\rangle = |1\rangle_f $. 
In other words the groundstate can be chosen as
$|GS\rangle = |0;0;\cdots ;0;\cdots \rangle_f $.

We can now see the influence of the $U_z$ operator previously defined on the groundstate.
One possible way to see this is to consider that when acting on a state with
$n_f=0$ the operator $-i\gamma_2 \gamma_3=I$, acts as the identity, $I_j$.
Acting with the unitary operator, $U_z$, and
using that $g^{\dagger}=(\gamma_1+i \gamma_2)/2$ and $i\gamma_1 \gamma_2=1-2g^{\dagger}g$, we find that
\bea
U_z |0\rangle_f &=& 
\prod_{j=1}^N \frac{1}{2} \left[ \gamma_1 +i\gamma_2 +\gamma_1 \gamma_3
-i \gamma_2 \gamma_3 \right]_j
|0\rangle_f \nonumber \\
&=& 
\prod_{j=1}^N \left[ 1+g^{\dagger}-g^{\dagger}g \right]_j |0\rangle_f
\eea
Using that 
$|\mathscr{S}_x=1\rangle \sim |\mathscr{S}_z=1\rangle + |\mathscr{S}_z=-1\rangle $
this implies that
$|0\rangle_f \sim |0\rangle_g + |1\rangle_g$.
So,
\be
U_z |GS\rangle \sim 
\prod_{j=1}^N \left[1+g^{\dagger} \right]_j |0\rangle_g \sim |\psi^+_0\rangle
\ee
with
$|\psi^+_0\rangle= |\psi^0_{even} \rangle + |\psi^0_{even} \rangle$, see Eq. \ref{a19}.
So, the states are
proportional, as expected \cite{greiter}. Using the unitary operator that transforms the spin Hamiltonian
to the Kitaev spinless fermion model we also transform between the groundstates 
of the two models:
\be
U_z |GS\rangle^{spins} \sim |GS\rangle^{Kitaev}_{g} \sim |GS\rangle_{d}
\ee

The operators $g$ and $f$ may be related by a unitary transformation that
transforms $\mathscr{S}_x$ into $\mathscr{S}_z$. Choosing a local
operator as
$\bar{U} = \frac{1}{\sqrt{2}} \left(1-\gamma_1 \gamma_3 \right)$,
we can show that
$\bar{U} f^{\dagger} \bar{U}^{\dagger} = i g $.
Its action on the spin operators may be determined and yields
\be
\bar{U} \mathscr{S}_x \bar{U}^{\dagger} = \mathscr{S}_z \hspace{1cm} 
\bar{U} \mathscr{S}_y \bar{U}^{\dagger} = \mathscr{S}_y \hspace{1cm}
\bar{U} \mathscr{S}_z \bar{U}^{\dagger} = -\mathscr{S}_x
\ee
Also,
\be
\bar{U} \gamma_1 \bar{U}^{\dagger} = \gamma_3 \hspace{1cm}
\bar{U} \gamma_2 \bar{U}^{\dagger} = \gamma_2 \hspace{1cm}
\bar{U} \gamma_3 \bar{U}^{\dagger} = -\gamma_1
\ee

The state $|0\rangle_f$ may be expanded as
$|0\rangle_f = \alpha |0\rangle_g + \beta |1\rangle_g$.
Acting with the creation operator $f^{\dagger}$ we may obtain that
$|1\rangle_f = -i \left(\alpha |1\rangle_g -\beta |0\rangle_g \right)$.
Imposing that $\langle 1|0\rangle_f=0$ we get $\beta^* \alpha=\alpha^* \beta$, and
using that $\langle \sigma_x=1| \sigma_z |\sigma_x=1\rangle =0$ we obtain that
$|\beta|^2=|\alpha|^2$.
Using that $|0\rangle_f$ is normalized $|\alpha|^2+|\beta|^2=1$
we get that $\alpha=\beta=1/\sqrt{2}$.

At zero temperature one may consider the operator $\sigma_x=-i \gamma_2 \gamma_3$ as the order
parameter of the ferromagnetic $XX$ spin chain. The average value of this operator in a state where
all the spins are oriented along the $x$ direction is different from zero. As we have seen the operator
may also be defined as $\sigma_x=- \left( 2 f^{\dagger} f -1 \right)$ and the groundstate
is defined by selecting the occupation numbers of the $f$ fermions as $n_f=f^{\dagger}f=0$ at
every lattice site. There is some liberty on calculating the average value of the order parameter
within the framework of the $XX$ chain. In addition to using the representation in terms of
the $f$ fermions we may use that
$\sigma_x=-i \gamma_2 \gamma_3 = -(g^{\dagger}-g) \gamma_3$.
Then the matrix element of the order parameter in the groundstate simplifies to
\be
\langle_f 0| \sigma_x | 0 \rangle_f = -\langle_f 0| \left( g^{\dagger}-g \right) \frac{f-f^{\dagger}}{i} |0\rangle_f
\ee
where we used that $\gamma_3=-i(f-f^{\dagger})$. Expanding the state $|0\rangle_f$ in terms of the states
$|0\rangle_g$ and $|1\rangle_g$ it is easy to show that
$\langle_f0|\sigma_x |0\rangle_f =1$.

This order parameter of the spin chain, for which there is a Landau type order and no topology,
has a dual in the topological Kitaev model. This can be obtained performing the same unitary
transformation, $U_z$, that is used to perform the duality transformation of the Hamiltonians.
The dual operator is therefore defined as
\be
U_z \sigma_{x,j} U_z^{\dagger} = U_z \left(-i \gamma_{2,j} \gamma_{3,j} \right) U_z^{\dagger}
\ee
Using the results of eq. (\ref{stringy}) we obtain that
\be
U_z \sigma_{x,j} U_z^{\dagger} = \left(-1 \right)^{\sum_{l=1}^{j-1} g_l^{\dagger} g_l}
\left(g_j^{\dagger}+g_j \right)
\ee

The eigenstates of the Kitaev model at the point that corresponds to the $XX$ chain ($\mu=0,t=\Delta$), 
are better expressed
in terms of the non-local operators $d_j,d_j^{\dagger}$ that diagonalize the Hamiltonian at this
point in parameter space (see Eq. (\ref{other}) in the Appendix A). We get therefore that
\be
U_z \sigma_{x,j} U_z^{\dagger} = \left(-1 \right)^{\sum_{l=1}^{j-1} g_l^{\dagger} g_l}
i \left(d_{j-1}^{\dagger}-d_{j-1} \right)
\ee
Using that
\be
\left( 1-2 g_{j-1}^{\dagger} g_{j-1} \right) \left( d_{j-1}^{\dagger} - d_{j-1} \right)
= -\left( d_j + d_j^{\dagger} \right)
\ee
we get that
\be
U_z \sigma_{x,j} U_z^{\dagger} = -i \left(-1 \right)^{\sum_{l=1}^{j-2} g_l^{\dagger} g_l}
\left(d_{j}^{\dagger}+d_{j} \right)
\ee
and therefore its average value vanishes. As discussed before \cite{greiter}, the dual operators
do not follow, in general, directly from standard order parameters even though
the procedure may provide interesting information on order parameters in topological
phases using as starting points order parameters in Landau like systems \cite{cobanera2}.
Recent work on possible order parameters in topological phases using the reduced density
matrix has been proposed \cite{wcyu}.

\subsection{Examples of other mappings}

Consider first the following unitary operator of class (v)
\be
U_v=\cos \frac{\theta}{2} \vec{n}_1 \cdot \vec{\gamma} + \sin \frac{\theta}{2} \vec{n}_2 \cdot \vec{\mathscr{S}}
= U_v^{\dagger}
\ee
with $\vec{n}_1 \cdot \vec{n}_2 =0$. Take for instance $\vec{n}_1=\vec{e}_x, \vec{n}_2=\vec{e}_y$.
Then
\be
U_v=\cos \frac{\theta}{2} \gamma_1 + \sin \frac{\theta}{2} \mathscr{S}_2 =
\cos \frac{\theta}{2} \gamma_1 -i \sin \frac{\theta}{2} \gamma_3 \gamma_1
\ee

The action of this operator on the $XY$ spin model choosing $\theta=\pi/4$ gives
\bea
U H_{XY} U^{\dagger} &=& \frac{i J_X}{2} \sum_j \left( \prod_{l=1}^{j-2} \left(-1 \right)^l \right)
S_{j,x} \gamma_{2,j+1} \nonumber \\
&-& \frac{i J_Y}{2} \sum_j \left( \prod_{l=1}^{j-2} \left(-1 \right)^l \right)
S_{j,y} \gamma_{1,j+1} 
\eea
Note that now there are cubic terms in Majorana operators and so the quartic
problem is not reduced to a quadratic problem. In other words this transformation
converts the Hamiltonian into a product of Majorana and spin operators, which, however, in
leading order does not conserve the fermionic number.

One may also consider the mapping along the same lines of an interacting spinfull fermion model such as
the Hubbard model to some effective spin/Majorana model. 
For each spin component one may introduce two Majorana operators and therefore there
are four Majorana operators in total for each lattice site. However, the spin operators
only require three Majorana operators and therefore one needs some sort of
enlargement of the Hilbert space such as by considering the extra Majorana operator
or by introducing an extra spin operator (a similar extension was considered before
\cite{nilsson3}). As a consequence some projection to the physical subspace is in general
required. This will be considered elsewhere.

\section{Non-local canonical mapping and topological invariants of the Kitaev model}

As mentioned above the original spin$-1/2$ problem is not topological while the
Kitaev model has topological phases \cite{kitaev}. 
We may determine the topological properties of Kitaev's model 
calculating the winding number \cite{yakovenko} or the Berry phase (Zak phase) \cite{niu} 
across the Brillouin zone.

\subsection{Winding number of fermionic problem}

\begin{figure}
\begin{center}
\includegraphics[width=0.65\columnwidth]{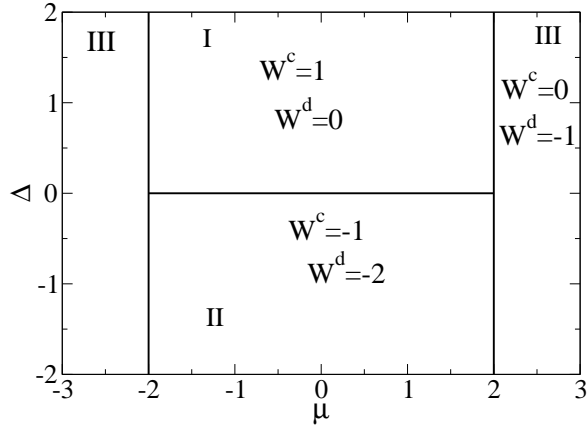}
\end{center}
\caption{\label{fig1}
Winding numbers $W^c$ and $W^d$ of the various phases as a function
of chemical potential and pairing amplitude in units of $t$. The winding number respecting
to the Hamiltonian in the representation of the $\tilde{d}$ operators takes the
values $W^{\tilde{d}}=0,-2,-1$ in phases $II,I,III$, respectively.
}
\end{figure}

In momentum space we may write the Kitaev model \cite{kitaev} as
\begin{eqnarray}
\hat H = \frac 1 2 \sum_k  \left( c_k^\dagger ,c_{-k}   \right)
\left(\begin{array}{cc}
\epsilon_k -\mu & -2 i \Delta \sin k \\
2 i \Delta \sin k & -\epsilon_k +\mu  \end{array}\right)
\left( \begin{array}{c}
c_{k} \\  c_{-k}^\dagger  \end{array}
\right)
\end{eqnarray}
with $\epsilon_k=-2t \cos k$. Here $c_k$ is the Fourier transform of $c_j$.
As is well known let us consider a chiral symmetry operator
$C$ such that
$C H C^{\dagger} =-H$.
In our case $C=\tau_x$, where $\tau_x$ is a Pauli matrix. Defining a matrix $T$ with columns
the eigenvectors of $C=\tau_x$ 
the winding number is defined as \cite{sato}
\be
W^c=\frac{1}{4\pi i} \int_{-\pi}^{\pi} dk \left[ q^{-1}
\frac{dq(k)}{dk} - (q^*)^{-1} \frac{dq^*(k)}{dk} \right]
\ee
where 
$q(k) = \epsilon_k-\mu+2i\Delta \sin k$ is the off-diagonal term of the hermitian matrix
$T H T^{\dagger}$ with null diagonal elements.
Calculating $W^c$ we get that in region $I$ of the phase diagram shown in Fig. \ref{fig1} $W^c=1$, in region
$II$ we get $W^c=-1$ and in the trivial phases we get $W^c=0$. As is well known \cite{kitaev},
$W^c$ counts the number of protected edge modes on each edge of the chain. Also
its sign depends on how one winds around the origin of the Brillouin zone.

Let us now consider the non-local canonical substitution eq. (\ref{other})
\be
c_j = \frac{i}{2} \left[ d_{j-1}^{\dagger}-d_{j-1} +d_j + d_j^{\dagger} \right]
\ee
These operators are specially useful at $\mu=0,\Delta=t$, but now we want to use them everywhere in the
phase diagram. Since it is non-local some care with boundary conditions has to be taken, but with
periodic boundary conditions the transformation is direct.
The Hamiltonian becomes more complicated. Kitaev's model in terms of these $d_j$ operators is
\bea
H &=& \frac{1}{2} \sum_j  (-t+\Delta) \left( -d_{j+1}^{\dagger} d_{j-1} -
d_{j-1}^{\dagger} d_{j+1} 
+d_{j-1} d_{j+1} + d_{j+1}^{\dagger} d_{j-1}^{\dagger} \right) \nonumber \\
&+& \frac{1}{2} \sum_j (t+\Delta) \left(2 d_j^{\dagger} d_j -1 \right)  \nonumber \\
&-& \frac{\mu}{2} \sum_j \left[ -d_{j-1}^{\dagger} d_j -d_j^{\dagger} d_{j-1}
+d_{j-1} d_j + d_j^{\dagger} d_{j-1}^{\dagger} \right] 
\eea

The chemical potential term now has nearest-neighbor terms both in the
hopping and pairing.
There is now a term with second-neighbors both in hopping and in pairing.
We get that the Hamiltonian in momentum space is given by
\begin{eqnarray}
\hat H = \frac 1 2 \sum_k  \left( d_k^\dagger ,d_{-k}   \right)
\left(\begin{array}{cc}
A & B \\
C & D
   \end{array}\right)
\left( \begin{array}{c}
d_{k} \\  d_{-k}^\dagger  \end{array}
\right)
\label{hkd}
\end{eqnarray}
with 
$A = \mu \cos k + (t+\Delta) +(t-\Delta)\cos 2k$,
$B =  i \mu \sin k +i (t-\Delta)  \sin 2k$, 
$C = -i \mu \sin k +i (-t+\Delta)  \sin 2k$, and
$D =  -\mu \cos k - (t+\Delta) -(t-\Delta)\cos 2k$.
This is of the type of the Kitaev model with second-neighbors.

The matrix $H_k^d$ from Eq. (\ref{hkd}) allows us to perform a chiral transformation which leads to
$q^d(k) = -(-t+\Delta) \cos 2k + (t+\Delta) + \mu \cos k
+ i (-t+\Delta) \sin 2k -i \mu \sin k$.
Consider for instance the special point $\mu=0,\Delta=t$. The winding number
\be
W^d=W^d(q^d(k)) = 0
\ee
So in terms of the $d$ operators the winding number in the special point indicates a trivial
phase. For instance at $\Delta=t, \mu>2t$ we get $W^d=1$. So it appears that from the
point of view of $W^d$ versus $W^c$ the model is ``dual" and an apparent change of topology
takes place.

Since in terms of the $d$-operators the Hamiltonian has first and
second neighbors, it is actually like an effective Kitaev model with first and second
neighbors. But in this case the model has a phase diagram which is richer with
winding numbers $W^d=0,\pm 1, \pm 2$. The model is in the BDI class with a $Z$ invariant.
For instance, considering $\mu=0$, we can show that it is equivalent to an effective
Kitaev model with parameters ($\mu_2,t_2,t_2^{\prime},\Delta_2,\Delta_e^{\prime}$)
where $\mu=0$ implies that $t_2=\Delta_2=0$ (no nearest-neighbor couplings, being hoppings or
pairings). It is well-known that this model has two Majoranas at each edge. With no nearest
neighbor couplings the model is like two decoupled chains and so the number of edge modes
just doubles. Solving the model in real space with open boundary conditions one obtains
two edge modes on each edge.

At point $\mu=0,\Delta=t$ we diagonalized the Kitaev
Hamiltonian in real space using Majorana operators and then introducing new fermionic
operators. We may achieve something similar at the point $\mu=0,\Delta=-t$. At this point
the fermionic operator that diagonalizes the Hamiltonian
may be defined as
$\tilde{d}_j = \frac{-i}{2} ( \gamma_{1,j} -i \gamma_{2,j+1} )$
This new operator allows to rewrite the Hamiltonian as
\be
H=t \sum_j \left( 2 \tilde{d}_j^{\dagger} \tilde{d}_j -1 \right)
\ee
the groundstate is obtained taking at each site the zero eigenstate of the operator $\tilde{d}_j^{\dagger} \tilde{d}_j$.
We have the relation 
\be
\tilde{d}_j = \frac{i}{2} \left[ c_j + c_j^{\dagger} + c_{j+1}^{\dagger} -c_{j+1} \right]
\ee
Replacing the operators
$c_j,c_j^{\dagger}$ by the operators $\tilde{d}_j,\tilde{d}_j^{\dagger}$ at an
arbitrary point in the phase diagram, we get a
similar expression just replacing $\Delta \rightarrow -\Delta$. 
We get that in the original
topological phase the winding number gives also zero.
In Fig. 1 we compare the winding numbers using the various representations
showing that their values depend on the set of operators used.

\subsection{Berry phase of fermionic problem}

Information about the topological properties may also be obtained calculating the Berry
phase associated with the eigenstates of the Hamiltonian. We may compare the Berry
phases using different representations
of the Hamiltonian of the system and therefore different states basis aiming for a more detailed
understanding of the relations between different fermionic representations of the problem.
We can show that
\begin{eqnarray}
\left( \begin{array}{c}
d_k \\ d_{-k}^{\dagger}
\end{array} \right) =
U_k^{\dagger}
\left( \begin{array}{c}
c_k \\ c_{-k}^{\dagger}
\end{array} \right),
U_k =
- e^{-\frac{ik}{2}} \left(\begin{array}{cc}
\sin \frac{k}{2}   & i \cos \frac{k}{2} \\
-i \cos \frac{k}{2}   & - \sin \frac{k}{2} \\
   \end{array}\right)
\label{similar}
\end{eqnarray}
which implies that
$H_k^d = U_k^{\dagger} H_k^c U_k$.
Performing the change from $c_k$ to $d_k$ corresponds to diagonalizing
the problem at $\mu=0,\Delta=t$ and performing the change from $c_k$ to 
$\tilde{d}_k$ corresponds to diagonalizing the problem at $\mu=0,\Delta=-t$.
At the points $\mu=0, \Delta=\pm t$ the eigenvalues are $\pm 2$
and the eigenvectors are
\begin{eqnarray}
\psi_+ =
sgn\left[\cos \frac{k}{2}\right] \left(\begin{array}{c}
-i \frac{\Delta}{t} \sin \frac{k}{2} \\
\cos \frac{k}{2}   
   \end{array}\right),
\psi_- =
sgn\left[\cos \frac{k}{2}\right] \left(\begin{array}{c}
\cos \frac{k}{2}   \\
-i \frac{\Delta}{t} \sin \frac{k}{2} 
   \end{array}\right)
\end{eqnarray}
Taking now $\mu=0$ but any value of $\Delta$, the eigenvalues are
\be
\lambda_{\pm} = \pm 2 \sqrt{ (t \cos k)^2 +(\Delta \sin k)^2}
\ee
The eigenvectors are
\begin{eqnarray}
\psi_+^{\Delta} =
\left(\begin{array}{c}
\frac{-i 2\Delta \sin k}{\sqrt{2 \lambda_+ \left( \lambda_++2t \cos k \right)}}
   \\
\sqrt{\frac{\lambda_+ +2 t\cos k}{2 \lambda_+}} 
   \end{array}\right), 
\psi_-^{\Delta} =
\left(\begin{array}{c}
\sqrt{\frac{\lambda_- -2 t\cos k}{2 \lambda_-}} 
   \\
\frac{-i 2\Delta \sin k}{\sqrt{2 \lambda_- \left( \lambda_- -2t \cos k \right)}}
   \end{array}\right)
\end{eqnarray}

Let us now consider the Berry phase in momentum space (Zak phase). For a given
eigenstate $n$ is given by
\be
\gamma_n = \frac{i}{\pi} \int dk \langle \psi_n(k) | \frac{\partial}{\partial k}
| \psi_n(k) \rangle
\ee
As we change variables (or operator descriptions) the states also change. Consider  
first one abelian change:
$|\psi_n(k)\rangle \rightarrow e^{i \xi(k)} |\psi_n(k)\rangle $.
Then we get that
\bea
\tilde{\gamma}_n &=& \frac{i}{\pi} \int dk \langle \psi_n(k) | e^{-i \xi(k)}
\frac{\partial}{\partial k} \left[ e^{i \xi(k)} |\psi_n(k) \rangle \right] \nonumber \\
&=& \frac{i}{\pi} \int dk \langle \psi_n(k) | \frac{\partial}{\partial k} | \psi_n(k)\rangle
-\frac{1}{\pi} \int dk \frac{d \xi(k)}{dk} \nonumber \\
&=& \gamma_n+\delta \gamma_n
\eea
If the function $\xi(k)$ is periodic the Zak phase is invariant. This is
similar to the well known case of adiabatic transport of some Hamiltonian that
depends on some parameter and one considers a cyclic transport: in this case the
Berry phase is invariant and observable and is related to the polarization of
a system of charges \cite{niu}.

In our case the state is a vector and in general we have a 
transformation from $|\psi_n\rangle$ to $|\tilde{\psi}_n\rangle$:
$|\tilde{\psi}_n\rangle = U^{\dagger} |\psi_n\rangle$,
and
$\tilde{H} = U^{\dagger} H U$.
Then we define
\bea
\gamma_n &=& 
\frac{i}{\pi} \int dk \langle \psi_n | \partial_k |\psi_n\rangle \nonumber \\
\tilde{\gamma}_n &=& \frac{i}{\pi} \int dk \langle \tilde{\psi}_n | 
\partial_k |\tilde{\psi}_n\rangle 
= \gamma_n + \delta \gamma_n 
\eea
where
\be
\delta \gamma_n = \frac{i}{\pi} \int dk \langle \psi_n |U \left( \partial_k U^{\dagger}
\right) |\psi_n\rangle
\ee

In general $\delta \gamma_n \neq 0$. We can see that
\bea
\gamma_n &=& \tilde{\gamma}_n +\frac{i}{\pi} \int dk \langle \tilde{\psi}_n |
U^{\dagger} \left( \partial_k U \right) |\tilde{\psi}_n\rangle 
\nonumber \\
\tilde{\gamma}_n &=& \gamma_n +\frac{i}{\pi} \int dk \langle \psi_n |
U \left( \partial_k U^{\dagger} \right) |\psi_n\rangle 
\eea
Defining a new phase as
\be
\tilde{\Gamma}_n = \frac{i}{\pi} \int dk \langle \tilde{\psi}_n |
D_k | \tilde{\psi}_n\rangle
\ee
with 
\be
D_k = \partial_k -\left( \partial_k U^{\dagger} \right) U
\ee
this new phase is invariant in the sense that $\Gamma_n=\tilde{\Gamma}_n$.
The differential operator $D_k$ is similar to a covariant derivative and similar
to a non-abelian gauge transformation (needed if states are degenerate, although
this is not the case here).

Let us now calculate the Zak phase of the state $|\psi_-\rangle_c$. This is given by
\bea
\gamma_{-1} &=& \frac{i}{\pi} \int_0^{2\pi} dk \langle \psi_- | \partial_k
|\psi_-\rangle_c \nonumber \\
&=& \frac{i}{\pi} \int_0^{2\pi} dk \frac{d}{dk} \left( \ln sgn \left[ \cos 
\frac{k}{2} \right] \right) \nonumber \\
&=& -1 = \gamma_c
\eea
Only the singular part of the wave functions contributes.

Calculate now the Zak phase of the lowest eigenstate of $H_k^d(\mu=0,\Delta=t)$.
This state is simply
\begin{eqnarray}
|\psi_-\rangle_d =
\left(\begin{array}{c}
0   \\ 1
   \end{array}\right)
\end{eqnarray}
and
$\gamma_{-1} = \gamma_d =0$.
Using that
$\gamma_d=\gamma_c + \delta \gamma$
and calculating $\delta \gamma$ we get $\delta \gamma=1$, showing that the change
of topology is hidden in the transformation.
We have checked that similar results occur in other topological models such as the topologically non-trivial
Shockley model.

\noindent
{\bf Singular vs. non-singular transformations}

Consider now a transformation between two points in parameter space. For instance,
two points at $\mu=0$ but with different values of $\Delta$.
Define diagonalized Hamiltonians
\bea
H_d^{\Delta} &=& U_{\Delta}^{\dagger} H_{\Delta} U_{\Delta}
\nonumber \\
H_d^{\Delta^{\prime}} &=& U_{\Delta^{\prime}}^{\dagger} H_{\Delta^{\prime}} 
U_{\Delta^{\prime}}
\eea
The eigenvalues are of the form
\be
\lambda_{\Delta} = 2 \sqrt{(t \cos k)^2+(\Delta \sin k)^2}
\ee
Then
$H_d^{\Delta^{\prime}}=(\lambda_{\Delta^{\prime}}/\lambda_{\Delta}) H_d^{\Delta}$ which
implies that
\be
H_{\Delta^{\prime}} = \frac{\lambda_{\Delta^{\prime}}}{\lambda_{\Delta}} \mathscr{U}^{\dagger}
H_{\Delta} \mathscr{U}
\ee
Here $\mathscr{U}=U_{\Delta} U_{\Delta^{\prime}}^{\dagger}$. We can relate the eigenstates
of the two Hamiltonians defined by
$H_{\Delta} |\psi_{\Delta}\rangle = \lambda_{\Delta} |\psi_{\Delta} \rangle $,
$H_{\Delta^{\prime}} |\psi_{\Delta^{\prime}}\rangle = \lambda_{\Delta^{\prime}} 
|\psi_{\Delta^{\prime}} \rangle $
as
$|\psi_{\Delta}\rangle = \mathscr{U} |\psi_{\Delta^{\prime}}\rangle$.

The Berry phases may be calculated as
\bea
\gamma_{\Delta} &=& \frac{i}{\pi} \int_0^{2\pi} dk \langle \psi_{\Delta} |
\partial_k |\psi_{\Delta} \rangle
\nonumber \\
\gamma_{\Delta^{\prime}} &=& \frac{i}{\pi} \int_0^{2\pi} dk \langle \psi_{\Delta^{\prime}} |
\partial_k |\psi_{\Delta^{\prime}} \rangle
\eea
As shown above they are related by
\be
\gamma_{\Delta^{\prime}}=\gamma_{\Delta}+
\frac{i}{\pi} \int_0^{2\pi} dk \langle \psi_{\Delta} |
\mathscr{U} \left(\partial_k \mathscr{U}^{\dagger} \right) |\psi_{\Delta} \rangle
\ee
Consider as an example, $\Delta=1,\Delta^{\prime}=-1$.
In these simple cases $\lambda_{\Delta}=\lambda_{\Delta^{\prime}}=1$.
The operator is simply given by
\begin{eqnarray}
\mathscr{U} =
\frac{1}{2}
\left(\begin{array}{cc}
2 \cos k    & 
2 i \sin k \\
2 i \sin k   & 2 \cos k  \\
   \end{array}\right)
\end{eqnarray}
Note that $1+\cos k \geq 0$ and no singular part appears. This suggests that
$\delta \gamma=0$. Indeed using that
\begin{eqnarray}
|\psi_-\rangle (\Delta=1) =
sgn \left[ \cos \frac{k}{2} \right]
\left(\begin{array}{c}
\cos \frac{k}{2}   \\ i \sin \frac{k}{2}
   \end{array}\right)
\end{eqnarray}
and
\begin{eqnarray}
|\psi_-\rangle (\Delta=-1) =
sgn \left[ \cos \frac{k}{2} \right]
\left(\begin{array}{c}
\cos \frac{k}{2}   \\ -i \sin \frac{k}{2}
   \end{array}\right)
\end{eqnarray}
we get that
$\gamma(\Delta=-1)=-1$, which is the same as for $\Delta=1$.

We may also change the parameters from a topological to a trivial phase.
For instance we may consider the point, $A$, $\mu=0,\Delta=t$ and the point, $B$,
$\mu>2t,\Delta=0$. The Hamiltonian at this trivial point is diagonal.
So the operator that diagonalizes is the identity. Therefore
$\mathscr{U}=U_{\Delta=t}$ and as we have seen before $U_{\Delta=1}$ is singular.
Therefore
$\gamma_B = \gamma_A+\delta \gamma=-1+1=0$. Note that the winding number
distinguishes the phases $I$ and $II$ in the phase diagram. It seems that the
Berry phase does not. However, $\gamma=-1=-\pi/\pi$ is the same as $\gamma=1$
since they differ by $2\pi/\pi$.

\section{Topology of spin model vs. fermionic representation}

The main point however is the change of topology as one transforms from
the spin problem to the fermionic dual problem. 
As established previously \cite{oshikawa2} the spin model is topologically trivial.
In general we can determine the topological properties of an interacting or 
non-interating system considering that the Hamiltonian depends on a variable
and consider that it is cyclic. For instance depends on some
angle like $H(\theta)$. The Berry phase is defined as
\be
i \gamma = \int_0^{2\pi} d\theta A(\theta)
\ee
where
$A(\theta) = \langle GS(\theta) | \partial_{\theta} | GS(\theta)\rangle$,
with $|GS(\theta)\rangle $ the groundstate obtained for a given value of
the cyclic parameter. 
One possible way to introduce a dependence on a cyclic variable is to
consider a local twist on a given link (this may be seen as twisted
boundary conditions \cite{thouless}) as
\be
S_i^+ S_j^- + S_i^- S_j^+ \rightarrow e^{i\theta} S_i^+ S_j^- + e^{-i\theta} S_i^- S_j^+
\ee
This method has been used \cite{hirano} to show that a spin-$1$ chain is topological while
half-integer spin chains are topologically trivial.
A difficulty arises 
when the Majorana representation is used 
since the spin problem is in general an interacting problem 
and some numerical exact diagonalization is required or
a Green's function approach may be used \cite{gurarie}.

Different conventional fermionic representations can be constructed from the Majorana
fermions. One possibility is to define a local transformation as
$f_j=(\gamma_{2,j}+i \gamma_{3,j})/2$ which implies that we can write
$i \gamma_{2,j} \gamma_{3,j}= 2 f_j^{\dagger} f_j -1$. In terms of these operators
we can write that
\be
H_X=\sum_j \left( f_j^{\dagger}f_j-\frac{1}{2} \right) \left( f_{j+1}^{\dagger} f_{j+1} -\frac{1}{2} \right)
\label{rep1}
\ee
This is an interacting fermionic problem and its topological properties may be
obtained imposing twisted boundary conditions and calculating the Berry
phase averaging over the twist angle \cite{thouless,wcyu}. However, in this representation, since
all terms are of the type of number density operators, there is no dependence on the twist angle and the
Berry phase vanishes, as expected of a topologically trivial system.
The same type of result was obtained before: the diagonalization of Kitaev's model
at the point $\mu=0,\Delta=t$ also allows to write the Hamiltonian in terms of the
density operator and therefore the topological invariant vanishes in the same manner.

Also, we may choose a non-local transformation of the type
$h_j=(\gamma_{2,j}+i \gamma_{3,j+1})/2$. This leads to
$i \gamma_{2,j} \gamma_{3,j+1} = 2 h_j^{\dagger}h_j-1$ and the Hamiltonian may be written as
\bea
H_X &=& \frac{1}{4} \sum_j \left( 2 h_j^{\dagger} h_j -1 \right) \left( h_{j-1} h_{j+1} +
h_{j-1} h_{j+1}^{\dagger} \right. \nonumber \\
&-& \left. h_{j-1}^{\dagger} h_{j+1} - h_{j-1}^{\dagger} h_{j+1}^{\dagger} \right)
\label{rep2}
\eea
which is of the type of correlated hoppings and pairings and that requires an
explicit calculation of the topological invariant. Note that the two sets of fermionic
operators are related by a transformation that is non-trivial and may lead to a change
of topology as above for the non-interacting (quadratic) problem. 
In this case the states are
not straightforwardly obtained and the change of the Berry phase involves now a trace over
a set of eigenstates that has to be obtained numerically. 
The relation between the operators $h$ and $f$ in momentum space is given by 
\begin{eqnarray}
\left( \begin{array}{c}
h_k \\ h_{-k}^{\dagger}
\end{array} \right) =
U_k^{\dagger}
\left( \begin{array}{c}
f_k \\ f_{-k}^{\dagger}
\end{array} \right) , 
U_k =
e^{-\frac{ik}{2}} \left(\begin{array}{cc}
\cos \frac{k}{2}   & i \sin \frac{k}{2} \\
i \sin \frac{k}{2}   &  \cos \frac{k}{2} \\
   \end{array}\right)
\end{eqnarray}
This raises the question of possible different topological
numbers of the interacting problem, depending on the fermionic representation
(see the similarity with Eq. \ref{similar}).

\begin{figure}
\begin{center}
\includegraphics[width=0.65\columnwidth]{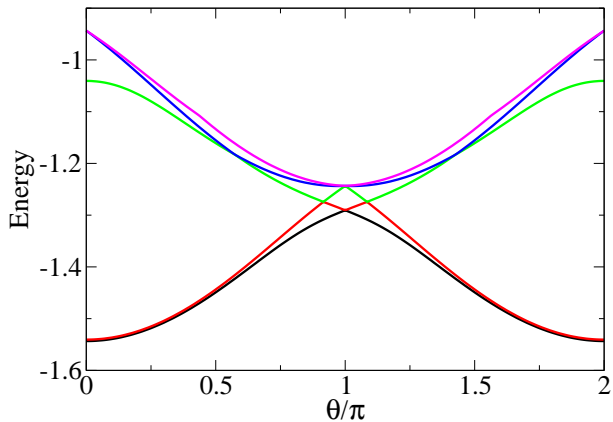}
\end{center}
\caption{\label{fig2}
Lowest energy levels obtained by exact diagonalization 
of a $N=8$ system of fermions that results from the Jordan-Wigner
transformation of a spin problem as a function of the angle of the twisted
boundary conditions, $\theta$. The parameters of the fermionic problem are
$t=1, \Delta=0.5$. ($4t=J_x+J_y$, $4\delta=J_x-J_y$).
}
\end{figure}

\noindent
{\bf Mapping and boundary conditions}

Let us consider again the spin problem
\be
H =-\sum_j \left( J_x S_j^x S_{j+1}^x + J_y S_j^y S_{j+1}^y \right)
\ee
The exact diagonalization of this Hamiltonian for a finite system shows that
using either periodic boundary conditions (PBC) or open boundary conditions (OBC) 
the groundstate is a singlet in general. For the $X$ model with $J_y=0$ the groundstate
is doubly degenerate for both sets of boundary conditions. Imposing twisted boundary
conditions one may calculate the Berry phase and it vanishes confirming that the
system is topologically trivial. 

Consider now the traditional Jordan-Wigner transformation
to spinless fermions. This is straightforward for OBC. However, as is well known,
choosing PBC for the spins, $S_{N+1}=S_1$, translates into the fermionic problem as
\be
c_{N+1}=e^{i \pi \hat{N}_F} c_1 = c_1 e^{i \pi (\hat{N}_F-\hat{n}_1)}
\ee
where $\hat{N}_F$ is the operator that counts the total number of fermions, $c$.
Diagonalizing the many-body fermionic problem that results from the Jordan-Wigner
transformation and considering OBC one obtains the same energy levels as for the
original spin problem, as expected. Implementing periodic boundary conditions
in addition to the relation imposed by the Jordan-Wigner transformation leads to
the same spectrum as for the spin problem with PBC. The exact diagonalization of the many-body
problem is carried out using an occupation number representation of the Hilbert space. 

Consider now the Kitaev model
by itself without reference to its spin origin. Diagonalization of the many-body problem
with OBC leads naturally to the same spectrum selecting the appropriate values of
$t=1.5,\Delta=0.5$ for $J_x=1,J_y=0.5$. Considering now the Kitaev model
with strict periodic boundary conditions $c_{N+1}=c_1$ leads however to
a different energy spectrum, in particular a different groundstate energy (at least for
a finite system).

Imposing twisted boundary conditions in the fermionic model that is obtained by the
Jordan-Wigner transformations leads to a Berry phase that vanishes, as for the original
spin problem. In Fig. \ref{fig2} we show the lowest many-body energy eigenvalues 
as a function of the twist angle, $\theta$, for $t=1, \Delta=0.5$. Note that the groundstate has no degeneracy
with the first excited state except perhaps at $\theta=\pi$.   

\begin{figure}
\begin{center}
\includegraphics[width=0.65\columnwidth]{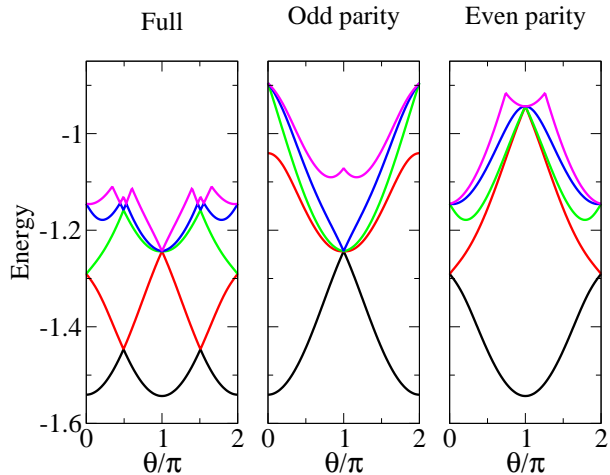}
\end{center}
\caption{\label{fig3}
Lowest energy levels of a $N=8$ system of fermions describing Kitaev's model 
as a function of the angle of the twisted
boundary conditions, $\theta$. The parameters of the fermionic problem are
$t=1, \Delta=0.5$. In the middle and right
panels a subset of states of given parity is considered. The Hamiltonian does
not couple the two subsectors.
}
\end{figure}

\begin{figure}
\begin{center}
\includegraphics[width=0.65\columnwidth]{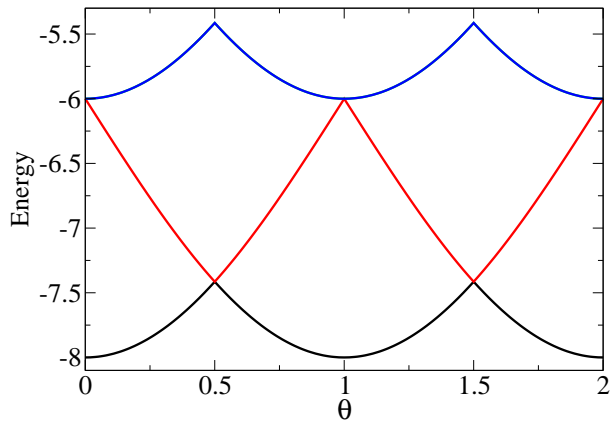}
\end{center}
\caption{\label{fig4}
Lowest energy levels of a $N=8$ system of fermions describing Kitaev's model 
as a function of the angle of the twisted
boundary conditions, $\theta$. The parameters of the fermionic problem are
$t=1, \Delta=1$. 
}
\end{figure}

Consider now the Kitaev model with strict PBC, once again solved as a many-body
problem. In Fig. \ref{fig3} and in Fig. \ref{fig4} we show
the lowest many-body eigenvalues as a function of the twist angle, $\theta$.
In Fig. \ref{fig3} we consider $t=1,\Delta=0.5$ and in Fig. \ref{fig4} we consider
$t=1,\Delta=1$. Note the degeneracy of the lowest energy state with the first
excited state at $\theta=\pi/2$. In Fig. \ref{fig3} we show in addition to the
full energy spectrum (left panel) the spectra for the two fermionic parities.
Since the Hamiltonian conserves the fermionic parity (since it only couples subspaces
where wither the number of fermions does not change or changes by two fermions) 
the Hamiltonian may be diagonalized by blocks. As shown before, the overall groundstate at
$\theta=0$ has odd parity. At $\theta=\pi/2$ the spectra cross and there is a degeneracy.
If one now calculates the Berry phase one finds that it is neither zero nor $\pi$ (mod($2\pi$).
The Berry phase is well defined if there is a gap throughout the twist angle space.

However, as is well known, and we have rederived above, using the single particle approach 
and using a momentum space description, the Berry phase is easily calculated.
We may then consider a real space description using twisted boundary conditions as for the full
many-body problem but considering the many-body states as Slater determinants of single-particle
states. In Appendix C the topological phase is confirmed.

In summary, the apparent change of topology, when the mapping
from a spin problem to the Kitaev model is carried out, is the result
of different boundary conditions (when OBC are not used) that lead to
a difference in the Berry phase with respect to the original spin problem.
The calculation of the Berry phase for the many-body fermionic Kitaev problem
is inconclusive due to degeneracies as a function of the twisted boundary conditions.
The analysis of the Berry phase of the dual interacting fermionic problems
that results from the representation of the spin operators by different fermionic
representations is further complicated by degeneracies introduced due to
the enlargement of the Hilbert space.

\section{Conclusions}

In this work we have considered a representation of fermionic and
spin operators in terms of Majorana fermions and have used it
systematically to relate the two types of problems.
We focused on mappings between the two types of systems with
particular emphasis on the relation of the topological properties
as a result of the unitary transformations that lead from one
problem to another and on the effect of boundary conditions. 
The simpler case of a unitary transformation that relates
two fermionic problems was considered and the singular or
non-singular nature of the transformation determines the change of
topological invariants. The analysis of the change of topological properties due to the 
mapping between the spin problem
and its fermionic representation is, however, more complex, since the
calculation of a topological invariant of the interacting problem
in general requires a numerical solution. This has been performed
using exact diagonalization of the system in the presence of twisted
boundary conditions. In the case of the mapping from the $XY$ model to
the Kitaev model using the appropriate boundary conditions in the fermionic
problem leads to a non-topological system (as the original spin problem)
while the Kitaev model has topological phases.

The various types of unitary transformations allow mappings to different
problems with preserved or changed statistics and in general allow
the replacement of interacting terms by free terms and vice-versa.
This property has been used in different contexts in the literature
to address the problem of strongly interacting systems even though in general
some enlargement of the Hilbert space occurs.
Another interesting problem may be the effect of interactions in the Kitaev model
and its possible mapping to spin problems \cite{katsura}.

\vspace{\baselineskip}
\vspace{\baselineskip}
\vspace{\baselineskip}

\noindent
{\bf Acknowledgments}

\vspace{\baselineskip}

PDS acknowledges several discussions with Stellan \"Ostlund, partial
support and hospitality by Henrik Johannesson and
by the Department of Physics of Gothenburg University grant 621-2014-5972 (Swedish Research Council), 
where the initial stages of this work
were carried out. The authors also acknowledge discussions with Bruno Amorim, Bruno Mera, Rubem Mondaini
and Yan Chao Li.

Partial support from FCT through grant UID/CTM/04540/2013
is acknowledged.

\vspace{\baselineskip}

\appendix

\section{Jordan-Wigner transformation and Kitaev chain}

We may consider a spin-$1/2$ system that has anisotropy in the $XY$ plane described by the Hamiltonian
\be
H_{XY} = -\sum_{j=1}^N \left( J_x S^x_j S^x_{j+1} + J_y S^y_j S^y_{j+1} \right)
\ee
Defining 
$S^x= \left( S^+ + S^- \right)/2; S^y=\left(S^+-S^- \right)/(2i)$
we can write that
\bea
H_{XY} &=& -\frac{1}{4} \sum_j \left[ \left( J_x-J_y \right) \left( S^+_j S^+_{j+1}
+ S^-_j S^-_{j+1} \right) \right] \nonumber \\
& & -\frac{1}{4} \sum_j \left[ \left( J_x+J_y \right) \left( S^+_j S^-_{j+1}
+ S^-_j S^+_{j+1} \right) \right] 
\eea
This is an interacting quantum problem that is well known to be diagonalizable.
A possible way consists in performing the transformation from the spin operators
to spinless fermionic operators as \cite{jordan}
\bea
S^+_j &=& c^{\dagger}_j e^{i \pi \sum_{n=1}^{j-1} c^{\dagger}_n c_n} \nonumber \\
S^-_j &=& e^{-i \pi \sum_{n=1}^{j-1} c^{\dagger}_n c_n} c_j 
\eea
leading to
\bea
H_{XY} &=& -\frac{1}{4} \sum_j \left[ \left( J_x-J_y \right) \left( 
c^{\dagger}_j c^{\dagger}_{j+1} + c_{j+1} c_j \right) \right] \nonumber \\
& & -\frac{1}{4} \sum_j \left[ \left( J_x+J_y \right) \left( 
c^{\dagger}_j c_{j+1} + c^{\dagger}_{j+1} c_j \right) \right] 
\eea

This model is related to the Kitaev model \cite{kitaev} (at vanishing chemical potential) if
we rewrite it as
\be
H = -t \sum_j \left( c^{\dagger}_j c_{j+1} + c^{\dagger}_{j+1} c_j \right)
 +\Delta \sum_j \left( c_j c_{j+1} + c^{\dagger}_{j+1} c^{\dagger}_j \right)
\label{eqa5}
\ee
choosing $t=(J_x+J_y)/4; \Delta=(J_x-J_y)/4$. Therefore if $J_y=0$ we get that
$\Delta=t$ and if $J_x=J_y$ we get that $\Delta=0$.
If $\Delta=0$ the spectrum is gapless and if $\Delta \neq 0$ there is a gap in the system.
Therefore, $\Delta=0$ is a critical point that separates two gapped phases.

Consider now the addition of a chemical potential term, $\mu \neq 0$.
This can be traced back to a magnetic field in the $XY$ model and adds a term of the type
$-\mu \sum_j \left( c^{\dagger}_j c_j -\frac{1}{2} \right)$.
For any $|\mu|>2t$ the band is either empty or full. For $|\mu|<2t$ the spectrum is gapless
if $\Delta=0$ and there is a gap if $\Delta$ is finite.

If $J_x,J_y>0$, the spin system is ferromagnetic. If $J_x>J_y$ ($J_x<J_y$) the spins order at zero
temperature along the $x$ direction ($y$ direction). If $J_x=J_y$ the system is isotropic and critical with power
law correlation functions. In the corresponding Kitaev model the critical regime is the
non-superconducting tight-binding model. If $J_x>J_y \rightarrow \Delta>0$ and $J_x<J_y \rightarrow \Delta<0$,
and in both cases there is a gap that corresponds in the spin problem to the Landau like quasi-long-range order.

The spinless fermionic operators may be written in terms of Hermitian operators,
Majorana operators, as
$c_j = \frac{1}{2} \left( \gamma_{1,j} + i \gamma_{2,j} \right)$.
In terms of these operators the Hamiltonian can be written as
\be
H = \frac{i}{2} \sum_j \left[ \left( -t+\Delta \right) \gamma_{1,j} \gamma_{2,j+1}
+ \left( t+\Delta \right) \gamma_{2,j} \gamma_{1,j+1} \right]
-\frac{i}{2} \mu \sum_j \gamma_{1,j} \gamma_{2,j}
\ee
Note that at some particular point $(\mu=0,t=\Delta)$ the Hamiltonian becomes
quite simple
\be
H=it \sum_j \gamma_{2,j} \gamma_{1,j+1}
\label{eq13}
\ee
If we define {\it non-local} operators \cite{kitaev}
$d_j^{\dagger} = \frac{1}{2} \left( \gamma_{2,j} - i \gamma_{1,j+1} \right)$
they are related to the other fermionic operators by
\be
d_j^{\dagger} = \frac{1}{2} \left[ -i (c_j - c_j^{\dagger}) -i (c_{j+1} +
c_{j+1}^{\dagger} ) \right]
\label{other}
\ee
We may note that
$2 d_j^{\dagger} d_j-1 = i \gamma_{2,j} \gamma_{1,j+1}$
which leads to
\be
H=t \sum_{j=1}^{N-1} \left( 2 d_j^{\dagger} d_j -1 \right)
\ee
If we use open boundary conditions (OBC) we see immediately that $d_N^{\dagger} d_N$ does not
appear in the Hamiltonian, which leads to a degenerate groundstate ($d_N^{\dagger} d_N=0,1$)
with two Majoranas.

If we use periodic boundary conditions (PBC) we just need to add this term ($d_N^{\dagger} d_N$):
\be
H=t \sum_{j=1}^N \left( 2d_j^{\dagger} d_j -1 \right)
\ee
Note that
$d_N = \frac{i}{2} ( c_N^{\dagger} -c_N + c_1 +c_1^{\dagger} ) $.
Therefore,
$d_N^{\dagger} + d_N = i ( c_N^{\dagger} -c_N )$ and
$d_N- d_N^{\dagger} = i ( c_1 +c_1^{\dagger} )$.
The groundstate of the $d$ operators is the groundstate of
the system, at this special point
\be
|GS \rangle = |n_d=0; n_d=0; \cdots ; n_d=0; \cdots \rangle
\ee
written in terms of the $d$ operators. 

We may now define two states with different fermionic parities \cite{greiter}
\bea
|\psi^0_{even} \rangle &=&  \left( \prod_{j=1}^{N-1} d_j d_j^{\dagger} \right) |0\rangle \nonumber \\
|\psi^0_{odd} \rangle &=&  \left( \prod_{j=1}^{N-1} d_j d_j^{\dagger} \right) c_N^{\dagger} |0\rangle 
\eea
where
$|0\rangle = |0\rangle_N \cdots |0\rangle_1$,
with
$c_j |0 \rangle_j =0$.
These states $|\psi^0_{even} \rangle $ and $|\psi^0_{odd} \rangle $ have no excitations of the $d$ operators.

These states can be represented in terms of the original
$c$ fermionic operators of the model as shown in Ref. \cite{greiter}
\bea
|\psi_0^{even}\rangle &=& \prod_{j=1}^N \left( 1+c_j^{\dagger} \right)_{even} |0\rangle_c
\nonumber \\
|\psi_0^{odd}\rangle &=& \prod_{j=1}^N \left( 1+c_j^{\dagger} \right)_{odd} |0\rangle_c
\label{a19}
\eea
where only even or odd powers of $c$ operators contribute. 
It was also shown that \cite{greiter}
\be
d_N |\psi_0^{odd}\rangle =0
\ee
and
\be
d_N^{\dagger} |\psi_0^{odd}\rangle = -i |\psi_0^{even}\rangle
\ee
One can conclude that the groundstate is non-degenerate and has odd parity.

\section{Generalized $SU(2)$ algebras and Majorana operators}

Considering the combinations $I_\pm=\frac{1}{2}(I\pm I_3)$, and
$\Sigma^\pm_i=\frac{1}{2}(\mathscr{S}_i\pm\gamma_i)$ one finds that the product of two operators of different signs vanishes and that 
\begin{eqnarray}
I_\pm^2&=&I_\pm \\
\Sigma^+_i\Sigma^+_j&=&\delta_{ij}I_++i\epsilon_{ijk}\Sigma^+_k \\
\Sigma^-_i\Sigma^-_j&=&\delta_{ij}I_-+i\epsilon_{ijk}\Sigma^-_k 
\end{eqnarray}
i.e. they commute and satisfy two separate $SU(2)$ algebras, with $I_\pm$ as the
identities. One also has
$(\vec{\Sigma}^\pm)^2=3I$ and $\vec{\Sigma}^+\cdot\vec{\Sigma}^-=0$.

Additionally, besides the projectors $I_\pm$, it is also sometimes useful to use the operators
$P^{+\pm}_{\vec{n}}=\frac{1}{2}(I_+\pm\vec{n}\cdot\vec{\Sigma^+})$ and
$P^{-\pm}_{\vec{n}}=\frac{1}{2}(I_-\pm\vec{n}\cdot\vec{\Sigma^-})$, which, for a
given unit vector $\vec{n}$, are also projectors, i.e. they satisfy $P_aP_b=\delta_{ab}P_a$, with
$a,b=\pm\pm$. 
Finite operators involving projectors are generally given by
\be
e^{i\alpha P}=I+(e^{i\alpha}-1)P=(I-P)+e^{i\alpha}P
\ee
for a single operator, and
\be
e^{i\sum_i\alpha_i P_i}=I+\sum_i(e^{i\alpha_i}-1)P_i
=(I-\sum_i P_i)+\sum_i e^{i\alpha_i}P_i
\ee
for several operators. When $\sum_i P_i=I$ the first term vanishes and one finds
the usual decomposition of an exponential operator in its own complete basis, namely
for the Boltzmann factor.
In particular, one has
\be
e^{i\alpha I_\pm}=I+(e^{i\alpha}-
1)I_\pm=I_\mp+e^{i\alpha} I_\pm
\label{eq_a}\ee

The finite operators of the $SU(2)$ generators are given by
\begin{eqnarray}
e^{i\frac{\theta}{2}\vec{n}\cdot\vec{\gamma}}&=&\cos\frac{\theta}{2}I+i\sin\frac{\theta}{2}\vec{n}\cdot\vec{\gamma}\\
e^{i\frac{\theta}{2}\vec{n}\cdot\vec{\mathscr{S}}}&=&\cos\frac{\theta}{2}I+i\sin\frac{\theta}{2}\vec{n}\cdot\vec{\mathscr{S}}\\
e^{i\frac{\theta}{2}\vec{n}\cdot\vec{\Sigma}_\pm}&=&I_\mp+\cos\frac{\theta}{2}I_\pm+i\sin\frac{\theta}{2}\vec{n}\cdot\vec{\Sigma}_\pm
\end{eqnarray}
with $\vec{n}$ a unit vector, $\vec{n}^2=1$. In the last equation, the first
term reflects the existence of two separate $SU(2)$ algebras, each with its own identity.
Finally,
\be
e^{i\frac{\theta}{2} I_3}=\cos\frac{\theta}{2}+i\sin\frac{\theta}{2}I_3
\ee
More general operators can be obtained from these results, in particular using the
last two equations and eq. (\ref{eq_a}), using
$\vec{n}\cdot\vec{\mathscr{S}}=\vec{n}\cdot\vec{\Sigma}_++\vec{n}\cdot\vec{\Sigma}_-$, 
$\vec{n}\cdot\vec{\gamma}=\vec{n}\cdot\vec{\Sigma}_+-\vec{n}\cdot\vec{\Sigma}_-$
and $I_3=I_+-I_-$

The action of finite operations, namely of unitary operations, on
the generators, are similar to the usual expression for the
rotations of vectors, but paying due attention to the existence of the two
$SU(2)$ commuting (and annihilating) algebras and to the parity of the different
operators in terms of the Majorana operators.

\section{Berry phase of Kitaev model: real space single-particle description}

The twisted boundary conditions for the wave functions of the Bogoliubov-de Gennes equations are taken as
\bea
u_n^{\theta}(j +N a) &=& e^{i \theta} u_n^{\theta} (j ) \nonumber \\
v_n^{\theta}(j +N a) &=& e^{-i \theta} v_n^{\theta} (j )
\eea
The problem is solved for a finite system with size $N$, typically taken large enough.
The groundstate may be represented by a matrix
\be
\Phi_{\theta}=
\left(\begin{array}{cccc}
\phi_{r_1}^{1,\theta} & \phi_{r_1}^{2,\theta} & \cdots & \phi_{r_1}^{M,\theta} \\
\phi_{r_2}^{1,\theta} & \phi_{r_2}^{2,\theta} & \cdots & \phi_{r_2}^{M,\theta} \\
\cdots & \cdots & \cdots & \cdots \\
\phi_{r_N}^{1,\theta} & \phi_{r_N}^{2,\theta} & \cdots & \phi_{r_N}^{M,\theta} \\
\end{array}\right)
\ee
where $M$ is the number of occupied single-particle states and $r_j$ is the coordinate of
site $j$. Also, $\phi_j^T = (u_j,v_j)^T$, where $T$ is the transpose of the vector.
The lattice Berry phase may then be obtained as
\be
i \gamma = \sum_{l=1}^L \ln A_l(\theta)
\ee
where $L$ is the number of $\theta$ points and
$A_l(\theta) = \langle GS(\theta_l) | GS(\theta_{l+1})\rangle $.
Since each state is now a many-body state given by a Slater determinant, the overlaps between two
states with different boundary conditions are given by
\be
\langle \Psi^{\theta} | \Psi^{\theta^{\prime}} \rangle
=det \left( \Phi_{\theta}^{\dagger} \Phi_{\theta^{\prime}} \right)
\ee
and the Berry phase is obtained by
\be
\gamma = -i \sum_{l=1}^{L} \ln \lambda_p
\ee
where $\lambda_p$ are the eigenvalues of the matrix product
$\left( \Phi_{\theta_l}^{\dagger} \Phi_{\theta_{l+1}} \right)$.
Considering a large enough system size and discretization of the twist angle between
zero and $2\pi$, leads to the expected result that the Berry phase is
$\pi$ in the topological regime previously identified and vanishes in
the topologically trivial region.

The degeneracy of the many-body spectrum at twist angle $\theta=\pi/2$
is understood looking at the single particle spectrum, at the same
twist angle. Indeed choosing $\theta=\pi/2$ leads to Majorana states
of vanishing energy at the edges of the system and therefore to the
degeneracy of the many-body spectrum. Interestingly this choice of
twisted boundary condition may be seen considering a ring pierced
by a flux of $\pi/2$. Note that the boundary condition is then
$c_{N+1}=\pm i c_1$.





\vspace{\baselineskip}

\noindent
{\bf References}

\vspace{\baselineskip}

\end{document}